\documentclass{appolb}
\usepackage{graphicx}
\usepackage{xcolor}
\usepackage{subfig}

\newcommand\sout{\bgroup\color{blue} \ULdepth=-.5ex \ULset}

\begin{document}
\title{Reconciling Multi-messenger Constraints with Chiral Symmetry Restoration \thanks{Presented at Quark Matter 2022}%
}

\author{Micha\l{} Marczenko\thanks{speaker}\thanks{e-mail: michal.marczenko@uwr.edu.pl}
\address{Incubator of Scientific Excellence - Centre for Simulations of Superdense Fluids, University of Wroc\l{}aw, plac Maksa Borna 9, 50-204 Wroc\l{}aw, Poland}
\\[3mm]
{Krzysztof Redlich, Chihiro Sasaki 
\address{Institute of Theoretical Physics, University of Wroc\l{}aw, plac Maksa Borna 9, 50-204 Wroc\l{}aw, Poland}
}
}
\maketitle
\begin{abstract}
We consider the parity doublet model for nucleonic and delta  matter  to investigate the structure of neutron stars. We show that it is possible to reconcile the multi-messenger astronomy constraints within a purely hadronic equation of state (EOS), which accounts for the self-consistent treatment of the chiral symmetry restoration in the baryonic sector. We demonstrate that the characteristics of the EOS required by the astrophysical constraints do not necessarily imply the existence of a hadron-quark phase transition in the stellar core. 
\end{abstract}
  
\section{Introduction}

The advancements of multi-messenger astronomy on different sources have led to remarkable improvements in constraining the equation of state (EoS) of dense, strongly interacting matter. The modern observatories for measuring masses and radii of compact objects, the gravitational wave interferometers of the LIGO-VIRGO Collaboration (LVC)~\cite{LIGOScientific:2018cki}, and the X-ray observatory Neutron Star Interior Composition Explorer (NICER) provide new powerful constraints on their mass-radius (M-R) profile~\cite{Riley:2019yda, Miller:2019cac, Riley:2021pdl, Miller:2021qha}. These stringent constraints allow for a detailed study of the neutron star (NS) properties and ultimately the microscopic properties of the EoS. In particular, the existence of $2~M_\odot$ NSs requires that the EoS must be stiff at intermediate to high densities to support them from gravitational collapse. At the same time, the tidal deformability (TD) constraint of a canonical $1.4~M_\odot$ NS from the GW170817 event implies that the EoS has to be fairly soft at intermediate densities, which may be indicative for a phase transition in the cores of NSs. This transition is commonly associated with a possible onset of deconfined quark matter. This conclusion has been achieved by systematic analyses of recent astrophysical observations within simplistic approaches (see, e.g.,~\cite{Alford:2013aca}). Although such schemes are instructive, they are not microscopic approaches. They provide interesting heuristic guidance, but cannot replace more realistic dynamical models for the EoS, which accounts for the fundamental properties of quantum chromodynamics (QCD), the theory of strong interactions, i.e., a self-consistent treatment of the chiral symmetry restoration in the baryonic sector. The recent LQCD results exhibit a clear manifestation of the parity doubling structure for the low-lying baryons around the chiral crossover~\cite{Aarts:2018glk}, Imprints of chiral symmetry restoration are also expected to occur in the baryonic sector of cold and dense matter. Such properties can be described in the framework of the parity doublet model~\cite{Detar:1988kn, Jido:2001nt}. The model has been applied to hot and dense hadronic matter, and neutron stars (see, e.g,~\cite{Marczenko:2021uaj,Zschiesche:2006zj, Marczenko:2017huu, Marczenko:2018jui, Sasaki:2010bp, Marczenko:2020jma, Marczenko:2020omo}).

In this work, we utilize the parity doublet model for nucleonic and $\Delta$ matter~\cite{Takeda:2017mrm} to investigate the implications on the structure of neutron stars.

\section{Equation of State}

The thermodynamic potential of the model in the mean-field approximation reads~\cite{Marczenko:2021uaj,Marczenko:2022hyt}
\begin{equation}\label{eq:thermo_potential}
	\Omega = V_\sigma + V_\omega + V_\rho + \sum_{x=N,\Delta}\Omega_x\rm,
\end{equation}
where $\Omega_x$ is the kinetic part of the thermodynamic potential, and $x$ labels positive-parity and negative-parity spin-$1/2$ nucleons, i.e., $N\in \lbrace p,n;p^\star,n^\star \rbrace$, and spin-$3/2$ $\Delta$'s, i.e., \mbox{$\Delta \in \lbrace\Delta_{++,+,0,-};\Delta^\star_{++,+,0,-}\rbrace$}. Note that the negative-parity states are marked with the asterisk. The potentials $V_i$ are commonly used mean-field potentials. The masses of the positive- and negative-parity chiral partners are given by
\begin{equation}\label{eq:doublet_mass}
	m^x_\pm = \frac{1}{2}\left[\sqrt{\left(g_1^x+g_2^x\right)^2\sigma^2 + 4\left(m_0^x\right)^2} \mp \left(g^x_1-g^x_2\right)\sigma\right] \textrm,
\end{equation}
where $\pm$ sign denotes parity and $x=N,\Delta$. When chiral symmetry is restored, the masses in each parity doublet become degenerate: $m_\pm^x(\sigma=0) = m_0^x$, where $m_0^x$ is the chirally invariant mass parameter. The positive-parity nucleons are identified as $N(938)$ states. Their negative-parity counterparts are identified as $N(1535)$ resonance~\cite{ParticleDataGroup:2020ssz}. The positive-parity $\Delta$ states are identified with $\Delta(1232)$ resonance. Their negative-parity chiral partners, $\Delta^\star$, are identified with $\Delta(1700)$ resonance~\cite{ParticleDataGroup:2020ssz}. Detailed description of the model and numerical values of the parameters used in this contribution can be found in~\cite{Marczenko:2022hyt}.

In the present work, we study the influence of $\Delta$ matter on the EoS and compliance with astrophysical constraints, i.e., $M_{\rm max} = (2.08 \pm 0.07)~M_\odot$~\cite{Fonseca:2021wxt}, as well as M-R and $\Lambda_{1.4} = 190^{+390}_{-120}$ from GW170817~\cite{LIGOScientific:2018cki}.

\section{Results}

\begin{figure}
    \centering
    \includegraphics[width=0.6\linewidth]{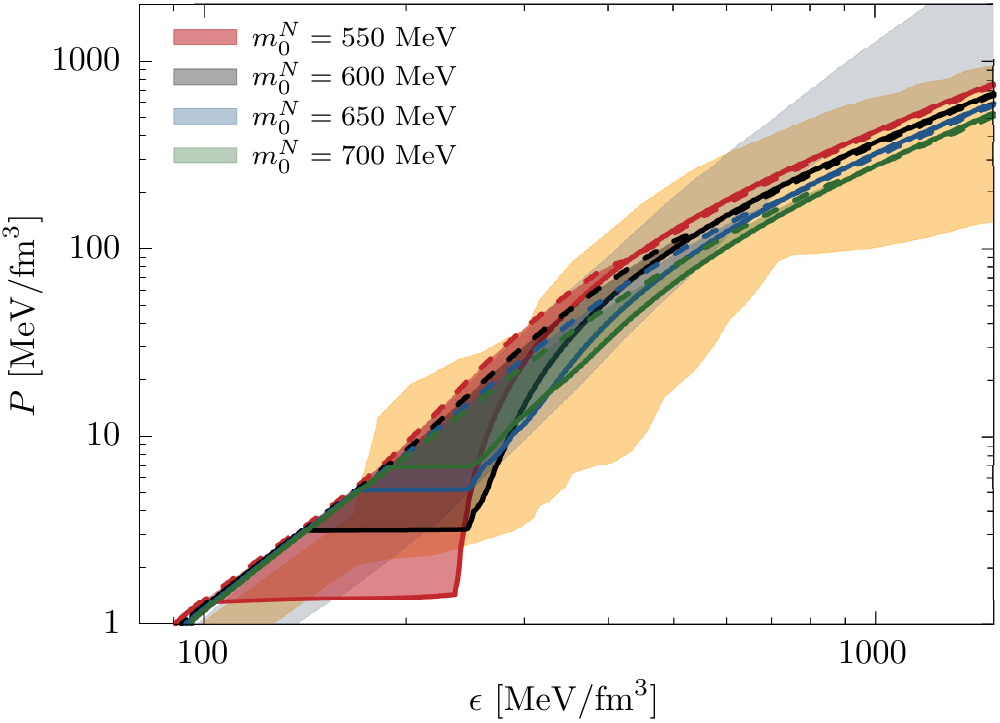}
    \caption{Thermodynamic pressure, $P$, under the NS conditions, as a function of the energy density, $\epsilon$. The dashed lines correspond to the purely nucleonic EoSs. The solid lines correspond to the case $m_0^N=m_0^\Delta$. The region spanned between the two lines marks the results obtained for $m_0^N < m_0^\Delta$. The region enclosed by solid (dashed) and dotted lines show solutions where $\Delta$ matter enters the EoS through a first-order (crossover) transition. The orange- and grey-shaded regions show the constraints obtained by~\cite{Annala:2019puf} and~\cite{LIGOScientific:2018cki}, respectively.}
    \label{fig:p_e}
\end{figure}

Fig.~\ref{fig:p_e} shows the calculated EoSs under the NS conditions for selected values of  $m_0^N=550$, $600$, $650$, $700~$MeV. To illustrate the effect of $\Delta$ matter on the EoS at intermediate densities, we show results obtained for purely nucleonic EoS (dashed line) together with the case $m_0^\Delta = m_0^N$ (solid line). The regions bounded by the two results correspond to the range spanned by solutions with $m_0^N < m_0^\Delta$ in each case. In general, the low-density behavior in each case is similar, until the deviations from the purely nucleonic EoSs are induced by the onset of $\Delta$ matter. The swift increase of the energy density is directly linked to the partial restoration of the chiral symmetry within the hadronic phase and resembled in the in-medium properties of dense matter in the parity doublet model. Most notably, it is associated with a drastic decrease of the mass of the negative-parity states in each parity doublet toward their asymptotic values, $m_0^x$. Interestingly, the softening is followed by a subsequent stiffening, as compared to the purely nucleonic result, and the EoS reaches back the constraints at higher densities. This effect is readily pronounced for $m_0^\Delta = m^N_0$. For other parametrizations shown in the figure, the EoSs fall into the region derived by the constraint. 

\begin{figure}
    \centering
    \includegraphics[width=0.6\linewidth]{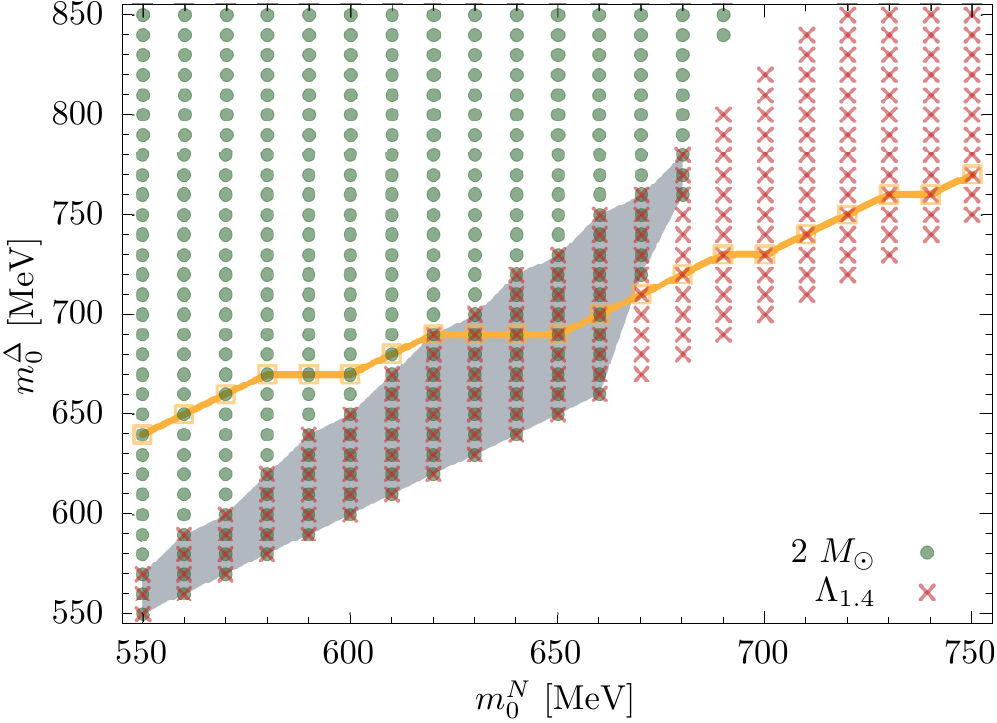}
    \caption{Allowed combinations of the model parameters, $m^N_0$ and $m^\Delta_0$.}
    \label{fig:m0_constraints}
\end{figure}

In Fig.~\ref{fig:m0_constraints}, we show the allowed combinations of $m^N_0$ and $m^\Delta_0$ for which the TD and $2~M_\odot$ constraints are met. The green circles indicate configurations that fulfill the lower bound for the maximum mass constraint, $M_{\rm max} = (2.08 \pm 0.07)~M_\odot$~\cite{Fonseca:2021wxt}. The red crosses indicate configurations that are in accordance with the upper bound for the TD constraint, $\Lambda_{1.4} = 190^{+390}_{-120}$~\cite{LIGOScientific:2018cki}. The gray-shaded area shows the region where the two constraints are fulfilled simultaneously. The orange points show configurations with the largest value of $m^\Delta_0$ for which the $\Delta$ matter appears through the first-order transition.

\begin{figure}
    \centering
    \includegraphics[width=0.6\linewidth]{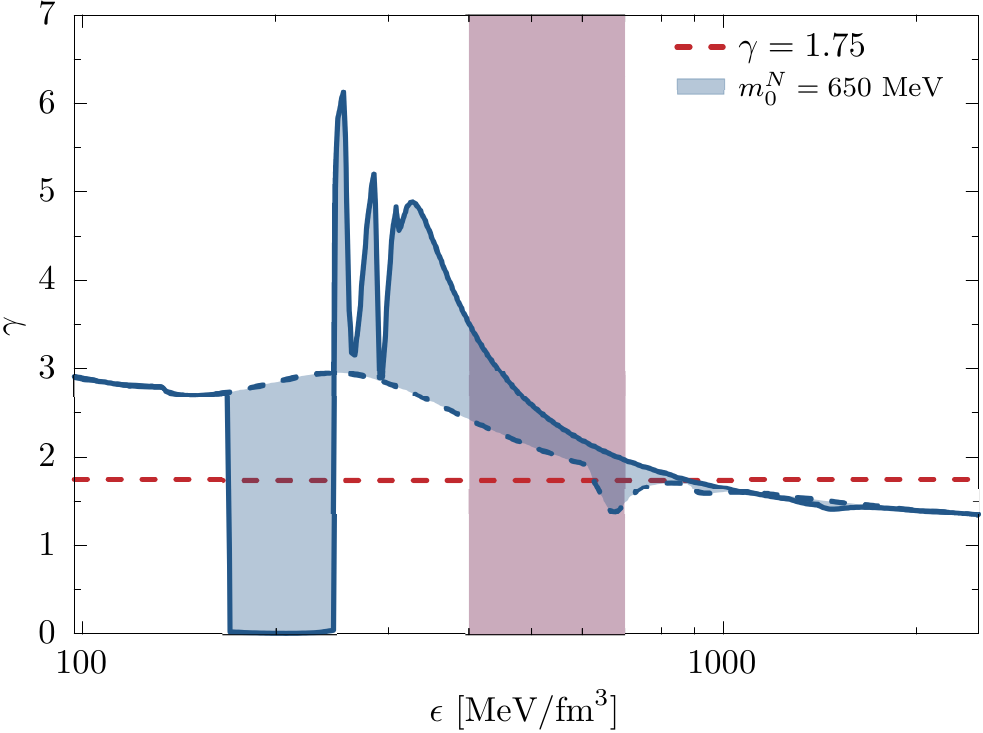}
    \caption{Polytropic index as function of energy density for $m_0=650~$MeV. Red, dashed line marks the threshold value for the onset of quark matter from Ref.~\cite{Annala:2019puf}. The purple band shows the energy-density range $\epsilon_{\rm QGP} = 400-700~$MeV for the onset of quark matter from lattice QCD~\cite{Bazavov:2014pvz}.}
    \label{fig:polytropic}
\end{figure}

The constraint derived in~\cite{Annala:2019puf} feature a notable change of the logarithmic slope of $p(\epsilon)$ around $\epsilon_{\rm QGP}\approx400-700~\rm MeV/fm^{3}$ (see Fig.~\ref{fig:p_e}), which is the estimate for the deconfinement transition at high temperatures~\cite{Bazavov:2014pvz}. It can be quantified by the polytropic index, i.e., $\gamma = d\log p / d\log \epsilon$. In~\cite{Annala:2019puf}, authors chose the criterion for the onset of quark matter in the core of NSs to be $\gamma < 1.75$. Interestingly, at higher densities, our results feature a similar change of the slope, regardless of the appearance of $\Delta$ matter. In Fig.~\ref{fig:polytropic}, we show as an example the polytropic index $\gamma$ obtained in the parity doublet model for $m_0^N=650~$MeV. Remarkably, $\gamma$ drops well below the threshold value of $1.75$ around $\epsilon_{\rm QGP}$. Thus, the polytropic index $\gamma$ does not provide a robust criterion and does not necessarily signal the onset of deconfined quark matter in the NS cores.

\section{Conclusion}

We have analyzed the properties of neutron stars and found that the multi-messenger constraints can be accommodated within a purely hadronic EoS for nucleonic matter including $\Delta(1232)$ resonance being subject to chiral symmetry restoration. As we have demonstrated in this work, the characteristics of the bulk EoS, such as the change of the logarithmic slope in the EoS, do not necessarily imply the existence of a hadron-quark phase transition as proposed in recent studies, e.g.,~\cite{Annala:2019puf}. We conclude that due to the anticipated near-future advances in multi-messenger astronomy, it will become inevitable to link the observed properties of NSs and their mergers to fundamental properties of strong interactions described by QCD, including chiral symmetry restoration, as well as emergence of conformal matter~\cite{Marczenko:2022jhl}.

\section*{Acknowledgements}
This work is supported partly by the Polish National Science Centre (NCN) under OPUS Grant No. 2018/31/B/ST2/01663 (K.R. and C.S.), Preludium Grant No. 2017/27/N/ST2/01973 (M.M.), and the program Excellence Initiative–Research University of the University of Wroc\l{}aw of the Ministry of Education and Science (M.M). K.R. also acknowledges the support of the Polish Ministry of Science and Higher Education.

\bibliographystyle{IEEEtran}  
\bibliography{main.bib} 

\end{document}